\documentclass[conference]{IEEEtran}
\IEEEoverridecommandlockouts
\usepackage{cite}
\usepackage{amsmath,amssymb,amsfonts}
\usepackage{algorithmic}
\usepackage{graphicx}
\usepackage{textcomp}
\usepackage{xcolor}
\usepackage{booktabs}
\usepackage{siunitx}
\usepackage{tabularx}
\usepackage{stfloats}

\usepackage[bookmarks=false, hyperfootnotes=false, pdfborder={0 0 0}, colorlinks=false]{hyperref}

\def\BibTeX{{\rm B\kern-.05em{\sc i\kern-.025em b}\kern-.08em
    T\kern-.1667em\lower.7ex\hbox{E}\kern-.125emX}}

\begin{document}

\title{Real-World Deployment and Performance Characterisation of Fog-Based Deep Learning for Cold-Chain Temperature Prediction over LoRaWAN\\

\thanks{ISBN 979-8-3195-1782-1/26/\$31.00~\textcopyright{}2026 IEEE}
\thanks{This work is based on the research supported in part by the National Research Foundation of South Africa (Ref Numbers SRUG22051310444, PMDS230703126166). The opinions, findings, conclusions or recommendations expressed in any publication generated by the NRF-supported research are those of the author(s) alone, and the NRF accepts no liability whatsoever in this regard.}
}

\author{

\IEEEauthorblockN{Jeremiah Taguta}
\IEEEauthorblockA{\textit{Department of Computer Science} \\
\textit{University of the Western Cape}\\
Cape Town, South Africa \\
0000-0001-8199-4307}
\and
\IEEEauthorblockN{Jean Frederic Isingizwe Nturambirwe}
\IEEEauthorblockA{\textit{eResearch Office} \\
\textit{University of the Western Cape}\\
Cape Town, South Africa \\
0000-0002-1794-7343}
\and
\IEEEauthorblockN{Clement Nthambazale Nyirenda}
\IEEEauthorblockA{\textit{eResearch Office} \& \\ \textit{Department of Computer Science} \\
\textit{University of the Western Cape}\\
Cape Town, South Africa \\
0000-0002-4181-0478}

}

\maketitle

\begin{abstract}

Fresh fruits and vegetables (FFVs) are highly perishable, and cold-chain breaks contribute significantly to global food waste. While Machine Learning (ML) can enable proactive intervention, cloud-based inference faces challenges such as latency and data loss. Fog computing addresses these issues but has been tested only in simulation for FFV cold-chain temperature prediction. To the best of the authors' knowledge, this paper presents its first real-world deployment. A fog-deployed LSTM-GRU model predicted cold-room temperature using LoRaWAN sensor data collected from a South African apple cold-storage facility with induced cold-chain breaks. Running entirely on a Raspberry Pi 4 with no cloud dependency, the system generated conditional SHAP explanations only when a break is predicted. The deployed system predicts cold-room temperature with an MAE of 0.2\textdegree{}C at roughly 0.2 kWh per day ($\approx$0.7 Wh per prediction). Predictions were delivered in under one second (555 ms), dominated by network and messaging rather than computation, with conditional explanations adding modest cost. SHAP consumes 28\% more CPU but is well within the hardware's capacity. The model attributes its predictions primarily to temperature, humidity, and their interaction. Critically, the deployment surfaced what simulation cannot: a sensor-triggered single point of failure, alongside genuine resilience, autonomous recovery from infrastructure faults and continued operation through internet loss. These are the first published deployment benchmarks for fog-based temperature prediction in FFV cold chains, establishing that explainable temperature forecasting is feasible on resource-constrained edge hardware. Future work includes asynchronous sensor fusion, commercial cold chain deployment, alternative model architectures, and causal analysis.

\end{abstract}


\section{Introduction}
Globally, a third of food produced for human consumption is lost or wasted across the supply chain, with fresh fruits and vegetables (FFVs) contributing nearly half of the loss, including in developing countries like South Africa \cite{noauthor_food_2017,goedhals-gerber_identifying_2024,taguta_towards_2026}. Because FFVs are highly perishable, proper cold chain monitoring is required because they are very sensitive to temperature fluctuations \cite{badia-melis_new_2018, goedhals-gerber_identifying_2024}, such that even short deviations result in temperature abuse, short shelf life, qualitative and quantitative losses, food insecurity, food-borne illness or death, financial loss and waste-driven global warming \cite{goedhals-gerber_identifying_2024, noauthor_food_2017, badia-melis_new_2018, taguta_fog-based_2026}. Food loss and waste also wastes resources such as water, energy and land \cite{noauthor_food_2017}. While the Internet of Things (IoT) has been utilised for real-time cold chain monitoring, it lacks predictive capability \cite{badia-melis_new_2018,vilas-boas_convergence_2023,taguta_towards_2026}. Machine learning (ML) offers proactive cold chain control by predicting temperature changes \cite{badia-melis_new_2018,j_taguta_comparative_2025}, but where ML models run from becomes a challenge. When deployed in the cloud, ML models may receive data late, delaying prediction and hindering proactive, real-time cold chain management \cite{solutions_unleash_2015}. The model may receive incomplete data, which can affect its accuracy and lead to unreliable predictions \cite{taguta_fog-based_2026}, potentially resulting in incorrect actions and increased FFV loss or waste. Late or inaccurate predictions are of no use; hence the need for low-latency ML inference, closer to the data source, reducing data loss, as offered by Fog computing \cite{solutions_unleash_2015}. It can therefore be argued that solving FFVs' cold chain monitoring challenges is beyond the capability of a single solution, but requires the integration of IoT, ML and Fog computing \cite{badia-melis_new_2018,vilas-boas_convergence_2023, taguta_towards_2026}. This study, therefore, investigates the integration of these technologies for increased FFV cold chain integrity and to minimise the massive global and South African burden of food loss and waste, and associated consequences.

While fog computing is well-suited to real-time applications, fog-based ML for FFV cold-chain monitoring remains underexplored \cite{taguta_towards_2026} and has, to date, been demonstrated only in simulation \cite{taguta_fog-based_2026}. To the best of the authors' knowledge, this study presents the first reported real-world deployment and measurement of an explainable ML (stacked LSTM-GRU (LSGR) architecture) temperature-prediction pipeline with multi-sensor fusion of independently timed (asynchronously arriving) sensors, combining the most recent reading from each at inference, on a fog node in an operational FFV cold room (a controlled cold room maintained at real cold-chain operating temperatures, with deliberately induced deviations). This study makes the following contributions:

\begin{itemize}

    \item The first published energy, pipeline latency, compute resource (CPU, RAM), and accuracy benchmarks for fog-based ML for temperature prediction in FFV cold-chains under real-world deployment.

    \item A demonstration that real-time LSGR inference and conditional SHAP explainability are viable on a Raspberry Pi 4 under sustained deployment, with explainability triggered only on temperature break predictions to minimise steady-state overhead.

    \item An event-driven sensor-fusion mechanism that integrates heterogeneous LoRaWAN sensors transmitting at the same nominal interval but with independent, unsynchronised timing, fusing the most recent available reading from each sensor at inference time.

    \item A characterisation of system fault tolerance under real field failures (battery depletion, power interruption, connectivity loss), demonstrating autonomous recovery without manual intervention.
\end{itemize}

The remainder of this paper is structured as follows. Section \ref{sec:related_work} reviews related work; Section \ref{sec:system} describes the proposed system; Section \ref{sec:setup} details the experimental setup; Section \ref{sec:results} presents and discusses the results; and Section \ref{sec:conclusion} concludes with key findings and future research directions.

\section{Related Work}
\label{sec:related_work}
Many studies have used ML to predict temperature in FFV cold chains. A Multilayer Perceptron (MLP) was utilised for predicting the temperature of apples during cold storage \cite{badia-melis_artificial_2016, loisel_machine_2022}, and citrus fruits and bananas during refrigerated transport \cite{zou_improved_2023}. Loisel et al. \cite{ loisel_machine_2022} also used Random Forest (RF), AdaBoost, Linear Regression (LR), Lasso, and Support Vector Machines (SVM). A 1D-Convolutional Neural Network predicted temperature at sensor-free locations in strawberry shipments \cite{ayanoglu_ml_2023}, while an Extreme Learning Machine (ELM), Decision Trees, LR, Naive Bayes, RF and SVM were compared for predicting the internal temperature of transport carrying tomatoes \cite{ali_iot-_2024}. A 4-layer LSTM predicted the internal temperature of navel oranges' cold rooms and the time to cold chain breaks \cite{guo_temperature_2024}. Moreover, Taguta et al.~\cite{j_taguta_comparative_2025} compared Gated Recurrent Unit, MLP, ELM and Extreme Gradient Boosting for apple cold room temperature prediction. Emenike et al.~\cite{emenike_c_c_improving_2016} predicted temperature in refrigerated transport carrying FFVs using Artificial Neural Networks. Of these studies, only two focused on the South African context \cite{emenike_c_c_improving_2016, j_taguta_comparative_2025}, and only one focused on the cold room phase and a single product \cite{j_taguta_comparative_2025}. Moreover, the reviewed studies do not explain or interpret the models, yet explainability is as important as model accuracy because it helps stakeholders understand how models arrive at decisions, increases trust, provides mechanisms to improve the models, and supports compliance reporting \cite{lundberg_unified_2017}. While these studies showcase the temperature-predictive power of ML, they utilised temporarily split existing datasets. Such validation, though useful, does not imply strong real-world deployment performance, as packet loss, stale data, asynchronous data transmission, and computing constraints characterise real-world deployment and require investigation \cite{taguta_towards_2026}.

Although Junjie et al.~\cite{junjie_environmental_2023} compared peephole LSTM and BP neural networks to predict temperature in a refrigerated transport vehicle carrying vegetables, inference runs on a personal computer receiving data from the cloud. Cloud-based prediction introduces latency unacceptable for real-time control. It depends on continuous, high-bandwidth connectivity that may be unavailable at farms, packing houses, and cold rooms in low-connectivity settings or refrigerated trucks driving through such settings \cite{ribeiro_junior_data_2022, musa_fog_2017,taguta_towards_2026}, where failed uploads mean no predictions and no proactive control \cite{musa_fog_2017, solutions_unleash_2015}. Fog computing brings computation and storage closer to the sensors \cite{musa_fog_2017}: a sensor and fog node connected over a local network guarantee low-latency inference without internet dependence \cite{ribeiro_junior_data_2022, taguta_towards_2026}, while data can be aggregated for later upload under reduced bandwidth when connectivity allows \cite{musa_fog_2017,ribeiro_junior_data_2022}. However, the reviewed studies did not use fog computing despite its benefits. A recent systematic review of ML for temperature prediction and temperature break detection and prediction in FFV cold chains confirmed the absence of fog computing and live inference on streaming sensor data \cite{taguta_towards_2026}. Even studies collecting data using IoT infrastructure used it as a sophisticated logging mechanism, training and evaluating models offline on the resulting datasets rather than during operation. Therefore, both the Fog–ML integration gap and the IoT–Fog–ML integration gap persist in research on FFV cold-chain temperature prediction. To the best of the authors' knowledge, Taguta et al.~\cite{taguta_fog-based_2026} was the first study to investigate the integration of IoT, Fog, and ML for predicting temperature in FFV cold chains using SimPy simulations. The prior work provides no mechanism to cost the model in terms of energy, RAM, CPU usage and latency on real-world deployment hardware, as the models were not deployed; a realism gap which remains unfilled. It lacks real-world characterisation of the deployment, especially on low-resourced fog nodes, yet real-world deployment validates the findings in the real operational world. The prior work did not study explainability, whose utilisation for FFV cold room temperature prediction models at the fog layer remains unexplored, yet critical for low-resourced hardware. This study, therefore, demonstrates the IoT-Fog-Explainable ML integration as a novel approach for FFV cold chain temperature prediction. Although the work in Taguta et al. \cite{taguta_fog-based_2026} validated this approach in simulation, the present study realises and characterises its deployment on real hardware, including on-device explainability.

\section{System and Method}
\label{sec:system}

\subsection{Architecture}
\label{sec:arch}
\begin{figure}[!ht]
  \centering
  \includegraphics[width=\columnwidth,clip,trim=0 0 0 0]{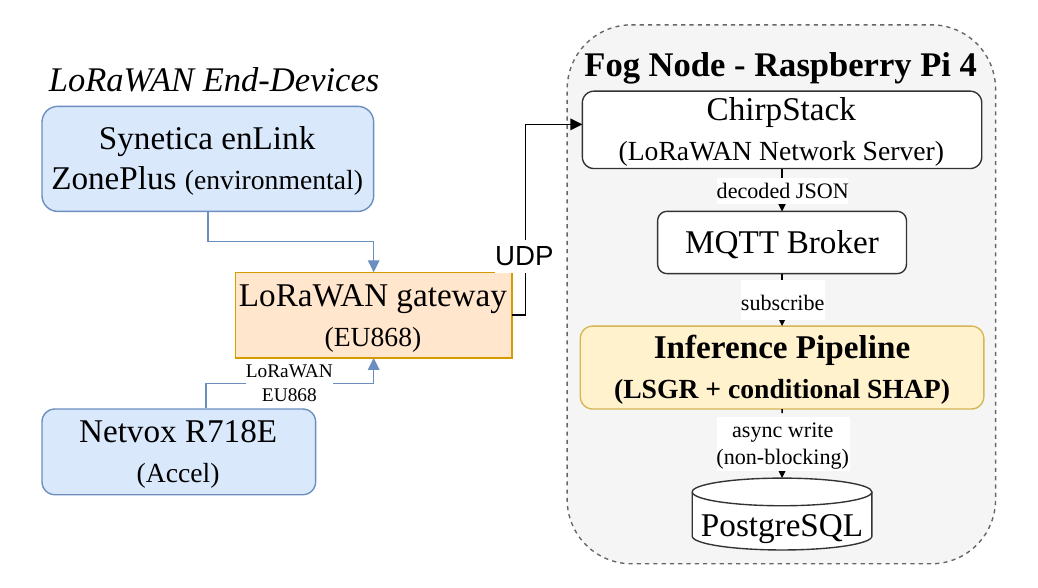}
  \caption{Fog-based architecture. All inference and explanation run on the fog node; no cloud services are used.}
  \label{fig:arch}
\end{figure}

The system is fog-centric: all sensing, inference, and explanation occur at the network edge, with no dependence on cloud infrastructure. Sensors transmit to a gateway over an EU868 LoRaWAN link, chosen for its low-power, long-range suitability for periodic environmental sensing \cite{taguta_towards_2026}. The gateway forwards uplinks to a single fog node that hosts the LoRaWAN network server, the inference pipeline, and the data store. The end-to-end path is \emph{sensors
$\rightarrow$ gateway $\rightarrow$ fog node (network server $\rightarrow$ broker $\rightarrow$ inference pipeline $\rightarrow$ database)}, as shown in Fig.~\ref{fig:arch}. Co-locating the network server, inference, and storage on the same node eliminates the need for the internet on the essential route, allowing the system to continue receiving data and predicting during internet failures, while also limiting operating costs to local energy rather than recurrent cloud charges.

\begin{figure}[!ht]
  \centering
  \includegraphics[width=0.8\columnwidth]{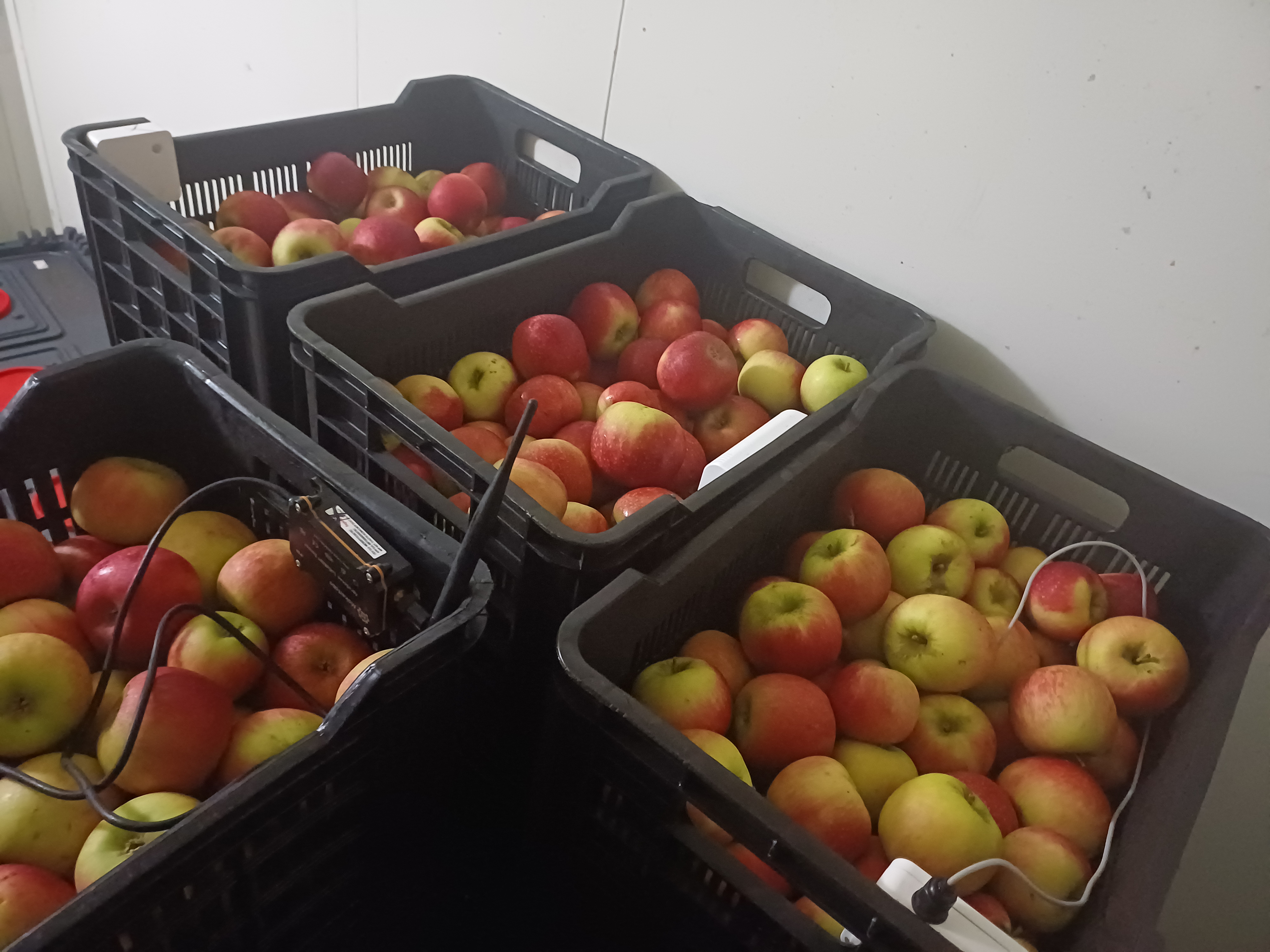}
  \caption{Cold room sensor deployment}
  \label{fig:experiment}
\end{figure}

Table~\ref{tab:hardware} presents a summary of the deployed hardware, while Fig.~\ref{fig:experiment} shows the sensor deployment. The gateway sends uplinks to ChirpStack (version 4.17.0) ~\cite{chirpstack}, running on a Raspberry Pi 4 fog node. ChirpStack decodes and publishes them to an Eclipse Message Queuing Telemetry Transport broker 2 \cite{junjie_environmental_2023,taguta_towards_2026}. A Python service subscribes, routes each message per device, and sends it to the inference pipeline. Background threads save decoded readings to PostgreSQL 14, ensuring that storage does not interfere with the inference path. 

\begin{table}[ht]
  \centering
  \caption{Deployed hardware.}
  \label{tab:hardware}
  \small
 \begin{tabularx}{\columnwidth}{@{}lXX@{}}
    \toprule
    \textbf{Component} & \textbf{Device} & \textbf{Role} \\
    \midrule
    Fog node & Raspberry Pi~4 & Edge server (full stack) \\
    Gateway  & Kerlink iFemtoCell-evo & LoRaWAN GW (EU868) \\
    Sensor   & Synetica enLink ZonePlus (eZone) & Environmental (primary) \\
    Sensor   & Netvox R718E & Vibration / acceleration (secondary) \\
    \bottomrule
  \end{tabularx}
\end{table}

The sensors communicate individually and asynchronously, with varying transmission patterns that include additional packets on motion events. Inference is event-driven: each valid reading from the environmental sensor updates a bounded rolling buffer and marks the system as ready to predict, and inference is triggered by the subsequent arrival of the vibration sensor's reading, which typically occurs a few seconds later, at which point the buffered environmental data and the vibration reading are fused into a single feature vector. Incomplete or motion-triggered packets that lack the necessary inference fields are identified and skipped, prohibiting predictions based on partial data. The system runs as managed services that start on boot and restart on failure; it therefore returns to a running state without manual intervention after power restoration or process failure, a behaviour examined in Section~\ref{sec:result_fault-tolerance}.

\subsection{Inference Pipeline and Conditional Explainability}
\label{sec:pipeline}

The predictive model is a stacked LSTM-GRU architecture (LSTM(64) - Dropout(0.5) - BatchNormalization – GRU(50)- BatchNormalization – Dense(1)), with both LSTM and GRU using tanh activation and sigmoid recurrent activation, trained using the Adam optimiser (learning rate: 0.0002), batch size of 32, and patience of 25. It was trained offline to predict cold-room temperature at the next timestep, using the dataset and feature-engineering approach described in Taguta et al. \cite{taguta_fog-based_2026}. The trained model and feature scaler are loaded once at startup and reused for all predictions, never reloaded per inference. Each inference cycle combines a rolling buffer of recent environmental readings with the latest vibration reading to form the engineered features (lagged, rolling-window, and interaction terms over temperature, humidity, CO\textsubscript{2}, and ambient light). The model predicts cold-room temperature at the next sampling interval, giving an observed median prediction horizon of 5 minutes. The predicted temperature is classified against configured thresholds: normal (within  \qty{-0.5}{\celsius} to \qty{1}{\celsius}) as required for the South African apples cold room \cite{anel_botes_ca_2019}, otherwise as a break.

Local feature attributions are computed conditionally using SHAP, preferred for robust local model interpretation \cite{lundberg_unified_2017}. SHAP's GradientExplainer was initialised with a background set of 100 records randomly sampled from the dataset described in Taguta et al. \cite{taguta_fog-based_2026}. The explanation was triggered only when a prediction was classified as a temperature break. During normal operation, no attribution is computed, and the explainability subsystem incurs no per-cycle costs. This conditional approach allows for on-device explainability while adhering to resource and latency restrictions. The cost of explanations is only incurred when they are operationally significant.

\section{Experimental Setup}
\label{sec:setup}

\subsection{Deployment}
\label{sec:deployment}
A Raspberry Pi 4B, 4-core ARM Cortex-A72 with 4GB RAM and 64GB storage, was used as a fog node. The system was deployed in an experimental or controlled Apple cold room in Stellenbosch, maintained at a nominal set-point of approximately \qty{-0.5}{\celsius} \cite{anel_botes_ca_2019}, and operated continuously from 25 May 2026. 54 crates of Sundowner apples ($\approx$100 per crate) were held in the cold room until 8 June 2026, when half were moved to a separate room for deviation injection; this study's sensors moved with them. Temperature breaks, or deviations, were induced to determine the model's performance under unstable, varying cold chain conditions. For this study, data up to 18 June 2026 (5619 prediction cycles) were utilised. All reported measurements were obtained under live operating conditions. The sensors transmitted at 5-minute intervals. During the deployment, the system experienced 5 key, naturally occurring field failures (battery depletion, power interruption, connectivity loss), which were used to characterise fault tolerance. Python 3.12.3 was used for programming. 

\subsection{Performance Metrics}
\label{sec:measurement}

\label{sec:setup-measurement}
To determine the energy, latency, and compute resource efficiency of the IoT-Fog-ML integration, the following metrics were used.

\subsubsection{Latency}
Latency was measured at successive pipeline stages from timestamps embedded at each hop. On-device stages are reported: network-server-to-broker (ns\_to\_mqqt) (from the server's receive time to wall-clock arrival, recorded as the first operation in the broker callback) and on-node processing (mqqt\_to\_pred) (broker arrival to completed inference), and SHAP latency, all three timed with Python's \texttt{$perf\_counter$}. mqqt\_to\_pred includes inference. 

\subsubsection{Energy}
Energy consumption was measured at the supply with an in-line smart-plug meter, recording energy over a representative 5-day period for the complete deployed system (fog node and gateway). Average per-day and per-prediction energy were derived from this total, the latter as total energy divided by the number of predictions over the 5 days.

\subsubsection{Compute resources}
CPU and memory usage were sampled in-process once per inference cycle. Memory is reported as the process resident set size sampled immediately after inference, that is, an instantaneous footprint of the inference pipeline. CPU is reported as average process utilisation over the inter-cycle interval; because it is a duty-cycle average, it reflects the mean load over the operation rather than the instantaneous load during inference.

\subsubsection{Model Accuracy}
The LSGR model's temperature predictions were evaluated using MAE, MSE (lower is better), and R\textsuperscript{2} (closer to 1 is better) \cite{j_taguta_comparative_2025}. Confidence intervals (CI) were obtained by bootstrap resampling (2 000 resamples)

\section{Results and Discussions}
\label{sec:results}
This section presents the results and discussion of this study, covering energy consumption, latency, compute resource usage, fault tolerance, model accuracy, and explainability.

\subsection{Energy consumption}
The system drew approximately 8 W on average, consuming about 0.2 kWh per day, equivalent to $\approx$0.7 Wh per prediction (amortised over 1457 predictions across the 5-day window). Total energy is dominated by the always-on hardware rather than by the light, intermittent prediction-and-explanation workload. This footprint, comparable to a low-wattage LED running continuously and far below server- or cloud-based inference, shows that explainable temperature prediction is feasible at very low energy cost on edge hardware.

\subsection{Latency decomposition}
\begin{figure}[ht]
  \centering
  \includegraphics[width=\columnwidth]{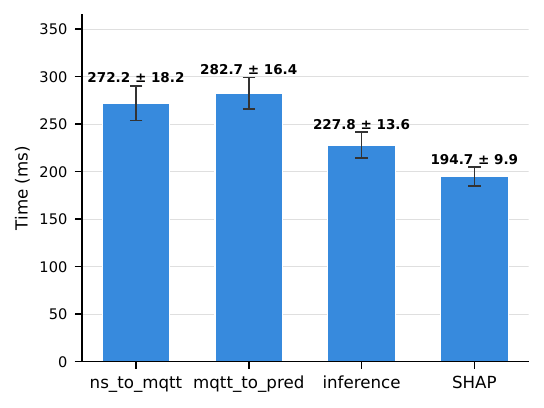}
  \caption{Decomposed latency.}
  \label{fig:result_latency}
\end{figure}

As shown in Fig.~\ref{fig:result_latency}, the per-stage latencies are consistent across runs, with low standard deviations. The ns\_to\_mqqt stage dominates, accounting for almost half of the latency up to inference, indicating that network-server processing and message passing, rather than on-device computation, are the principal costs. On-node processing, including preprocessing and inference of the uncompressed model, takes approximately 283 ms (228 ms for inference); this is modest given that the model runs without compression on edge hardware. When a break is predicted, conditional SHAP adds approximately 195 ms to attribute features.

The system produces a prediction in about 555 ms (excluding the gateway-to-network-server hop). While significant for hard real-time systems, this remains well under one second, a small fraction of the 300 s inter-reading interval, and is therefore near real-time for cold-chain monitoring; even with explanation, the total stays below one second. Notably, the cost of explainability is lower than that of inference, an acceptable price for transparency. The measured sensor-to-prediction latency is substantially higher than the 84.31 ms reported in the simulation study \cite{taguta_fog-based_2026}. This reflects the gap between simulation and real deployment: the simulation abstracted execution and data transmission as idealised costs, whereas the hardware realisation incurs their true overheads, which are framework execution, messaging, and physical LoRaWAN transmission. The discrepancy underscores the value of deployment measurement, as these realities are concealed by abstraction yet dominate real-world latency.

\subsection{Compute resources}
Table \ref{tab:results_resources} shows that the memory cost of inference is consistently less than 20\%. This implies that the Pi 4 has enough memory for temperature prediction tasks using DL models. 

\begin{table}[!h]
  \centering
  \caption{Resources efficiency.}
  \label{tab:results_resources}
 \begin{tabularx}{\columnwidth}{@{}lXXXX@{}}
   
    \toprule
     & CPU util All (\%) &CPU util no SHAP (\%) & CPU util SHAP (\%)& RAM All (MiB) \\
    \midrule
    Mean 	& 1.95 $\pm$0.62  &1.89 $\pm$0.51  & 2.42 $\pm$1.06   & 727.41 $\pm$6.51 \\
    Median	                 & 1.70          & 1.70  & 1.90 & 724.90 \\
    Q3 		                 & 2.00          & 1.90   & 3.20 & 730.50 \\
    
    \bottomrule
    
  \end{tabularx}
\end{table}
\noindent The table also shows that CPU usage spiked and was inconsistent, with an average of 1.95\%, a third quartile of 2\% and a maximum of 10.7\%. The CPU burst indicates events which are triggered, and this is confirmed by CPU utilisation without SHAP upon predicting normal temperature (mean: 1.89\%), which is consistently low, compared to with SHAP (after break prediction) (mean: 2.42\%), marking a 28\% CPU usage increase, even with a higher utilisation median. The utilisation spiked due to SHAP computations. This implies that SHAP increases CPU utilisation and is computationally expensive, though the Pi 4 still maintained a low load and handled the computations. Since RAM usage was measured after inference but before SHAP, no significant differences were visible between SHAP and no SHAP cycles. The low CPU utilisation and the Raspberry Pi's ability to sustain the explainable ML pipeline demonstrate that a single fog node is computationally sufficient for this workload, providing real-world evidence for the feasibility previously found in simulation \cite{taguta_fog-based_2026}. CPU utilisation remained low despite inference and explanation taking over 200 ms (Fig.~\ref{fig:result_latency}). This reflects that the latency is largely attributable to framework and interpreter overhead rather than sustained computation, and that the brief processing burst is averaged out over the longer CPU-sampling window; that is, the workload is latency-bound rather than compute-bound.

\subsection{Fault tolerance and resilience}
\label{sec:result_fault-tolerance}

Evaluating the system in a real deployment, rather than in simulation, exposes failure modes that arise only under genuine operating conditions. Over 24 operational days, the system experienced 5 interruptions of 1.5~hours or more, totalling 79.1~hours of downtime (no predictions) and corresponding to 86.31\% availability; Table~\ref{tab:results_faults} lists each. Most required manual intervention to fix the problem, though the service automatically resumed after the fix.

\begin{table}[!ht]
\centering
\caption{Fault categories observed during the deployment.}
\label{tab:results_faults}
\footnotesize
\begin{tabularx}{\columnwidth}{p{2cm} p{1.1cm} X X}
\toprule
\textbf{Cause} & \textbf{Duration} & \textbf{Recovery} & \textbf{Data impact} \\
\midrule

eZone battery depletion &
16.9 ~h &
Manual battery replacement  &
Inference halted and automatically resumed \\

eZone battery low &
1.5 ~h &
Automatically woke up  &
Inference halted and automatically resumed  \\

Connectivity loss and loss of the external (Internet) link &
22.4~h &
None required; local operation unaffected &
Local inference and logging continued; only the remote access was affected \\

Failure of the Raspberry Pi power cable &
12~h &
Manual replacement &
Inference paused, then resumed automatically\\

IP reassignment disrupted the gateway's packet forwarding. &
48.7~h &
Static IP assigned to the host &
Inference halted, and automatically resumed \\
\bottomrule
\end{tabularx}
\end{table}

These results characterise reliability without claiming fault-free operation. The system recovered autonomously from infrastructure failures and continued local inference during internet loss, confirming its cloud-independent operation, which is desirable for cold-chain environments where connectivity may be intermittent, while also revealing a sensor-level vulnerability. Reporting both is essential to an honest account of real-world deployment, and the observed failure modes provide design guidance that simulation-based evaluation could not. However, a structural limitation of the deployed design is that inference is triggered by the arrival of the Netvox supplementary readings, creating a single point of failure: if that sensor becomes unavailable, inference halts even if the environmental sensor remains active. Future work will develop a redesigned trigger that proceeds from the environmental sensor and degrades gracefully.

\subsection{Explainable Temperature Prediction}

Predicting one sampling interval ahead, the LSGR model achieved an MAE of \qty{0.1972}{\celsius} (95\% CI [0.1834, 0.2146]), an MSE of 0.3826 (95\% CI [0.1519, 0.6874]), and an R\textsuperscript{2} of 0.7683 (95\% CI [0.5846, 0.8961]) under real-world deployment conditions. The low MAE indicates that predictions track actual temperature closely under typical operation and are highly operationally accurate, while the higher MSE and moderate R\textsuperscript{2} reflect the model's difficulty anticipating sharp, genuine spikes in advance, a real-world challenge left to future work. The model exhibits a small positive bias (\qty{+0.13}{\celsius}), tending to slightly over-predict temperature. In cold-chain monitoring, where breaks manifest as rising temperature, this is the fail-safe direction: a model that errs slightly warm is biased toward flagging deviations rather than missing them. This implies the model is generally hot, or biased toward hotter temperature predictions. 

\begin{figure}[!ht]
  \centering
  \includegraphics[width=\columnwidth]{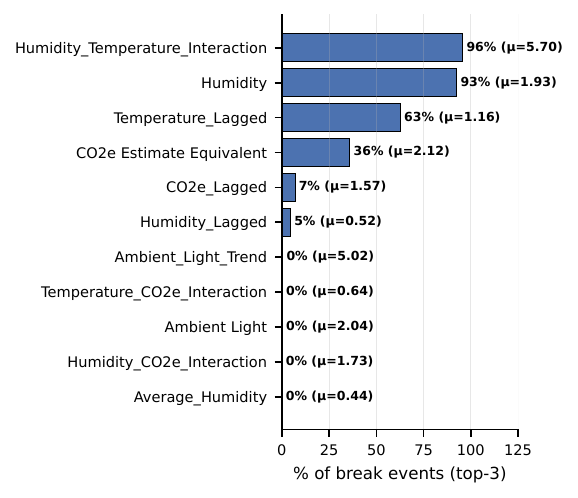}
  \caption{$\mathit{Bar}$ = frequency as top-3 driver; $\mu$ = mean $|\text{SHAP}|$ when present ($n$ = 631 break events)}
  \label{fig:shap_freq_with_magnitude}
\end{figure}

Fig.~\ref{fig:shap_freq_with_magnitude} shows that humidity and the single temperature lag appeared most frequently in the top three important features, indicating the model relies on them most for break prediction; the prominence of the humidity–temperature interaction suggests their combination is a major contributor. In contrast, CO\textsubscript{2}, ambient light, and acceleration ranked consistently low, contributing little to the model's predictions over the observed conditions. As the features the model relies on the most, these are the quantities most worth monitoring in operation, offering practical insight into developing deviations regardless of whether they are causal drivers. Establishing their causal role would require dedicated causal analysis and controlled intervention, which is left to future work. Since predictions rely on a small feature subset, these also form a candidate reduced feature set for future model improvement.

\subsection{Limitations of the study}
Tying predictions to both sensors causes system failure, especially when the Netvox provides only supplementary data. The study also uses a controlled laboratory cold room, which excludes operational routines like regularly loading and removing products. These activities introduce door openings and airflow disruption that increase the variance of cold-room conditions, making the prediction task more demanding than the one evaluated here; performance under commercial handling regimes may differ. Breaks were also induced deliberately rather than arising from equipment or handling faults, so the findings were not obtained under commercial operating conditions. The radio-and-gateway hop was omitted, as differencing gateway and network-server clocks produced some negative values and outliers, indicating clock offset between the two devices.

\section{Conclusion \& Future Work}
\label{sec:conclusion}
This paper presented the real-world deployment and characterisation of an explainable, fog-based LSGR pipeline for FFV cold-chain temperature prediction, realising on physical hardware a system previously validated only in simulation. Deployed in a real cold room under operational conditions, the pipeline achieved high operational accuracy (MAE) but was slightly affected by spikes while running entirely on a fog node, with low energy per prediction, reasonable latency, low resource cost, and on-device explainability at no steady-state overhead. SHAP attributed predictions primarily to past temperature, humidity and their interaction, and the Raspberry Pi 4 proved operationally sufficient for explainable temperature prediction. Beyond these benchmarks, the deployment surfaced realities that simulation cannot: autonomous recovery from infrastructure faults, continued operation through internet connectivity loss, and a single point of failure arising from the sensor-triggered inference design. Future work will redesign the inference trigger so prediction degrades gracefully when a sensor is unavailable, quantify predictive lead time ahead of breaks, explore adaptive learning approaches to improve accuracy, investigate on-device causal inference, and extend evaluation to commercial cold-chain conditions, other models, commodities and cross-device timing. Overall, the study demonstrates that a single fog node can deliver practical, autonomous, near-real-time explainable cold-chain temperature prediction for low-connectivity sites at an affordable cost, with feature attributions identifying which variables warrant closest monitoring and pointing toward future model simplification.

\bibliographystyle{IEEEtran}
\bibliography{Library}

\end{document}